\documentclass{aastex}
\include{epsf}
\usepackage{emulateapj5}

\def\Msun{\hbox{$\rm\thinspace M_{\odot}$}}

\def\h50{\hbox{$\rm\thinspace h_{50}$}}
\def\h50m1{\hbox{$\rm\thinspace h_{50}^{-1}$}}

\def\P3M{\hbox{$P^{3}M$}}

\def\AP3M{\hbox{$AdP^{3}M$}}

\def\cc2{c2}
\def\cc3{c3}
\def\cc4{c4}
\def\cc{c}

\def\etal{{\it et al.\thinspace}}

\def\spose#1{\hbox to 0pt{#1\hss}}
\def\approxlt{\mathrel{\spose{\lower 3pt\hbox{$\sim$}}
	\raise 2.0pt\hbox{$<$}}}
\def\approxgt{\mathrel{\spose{\lower 3pt\hbox{$\sim$}}
	\raise 2.0pt\hbox{$>$}}}

\def\<{\thinspace}

\def\boxit#1{\vbox{\hrule\hbox{\vrule\kern3pt\vbox{\kern3pt
          #1 \kern3pt}\kern3pt\vrule}\hrule}}

\shorttitle{The Metallicity of Pre-Galactic Globular Clusters}
\shortauthors{Beasley et al.}

\begin{document}

\title{The Metallicity of Pre-Galactic Globular Clusters:
Observational consequences of the first stars} 

\author{Michael A. Beasley\altaffilmark{1}, Daisuke Kawata\altaffilmark{1},
Frazer. R. Pearce\altaffilmark{2}, Duncan A. Forbes\altaffilmark{1} \& Brad K. Gibson\altaffilmark{1}}

\altaffiltext{1}{Centre for Astrophysics and Supercomputing,
 Swinburne University of Technology, Hawthorn, VIC 8122, Australia 
\email{mbeasley,dkawata,dforbes,bgibson@astro.swin.edu.au}}
\altaffiltext{2}{Department of Physics and Astronomy, University 
of Nottingham, Nottingham NG7 2RD, UK.
\email{Frazer.Pearce@nottingham.ac.uk}}

\begin{abstract}

We explore a scenario where metal-poor globular clusters (GCs) 
are enriched by the first supernovae in the Universe. 
If the first stars in a 10$^7$ M$_\odot$ dark halo were 
very massive ($>180$ M$_\odot$), then a pair instability supernova from
a single massive star can produce sufficient iron to enrich $10^6\Msun$
of pristine, primordial gas to ${\rm [Fe/H]} \sim -2$.
In such a scenario, where a single massive star acts as a seed for 
halo GCs, the accurate abundance analysis of GC stars would allow
a direct measurement of the Population III initial mass.
Using the latest theoretical yields for zero metallicity stars
in the mass range 140$-$260 M$_\odot$, 
we find that the metals expelled from a $\sim230$ M$_\sun$ star are 
consistent with [Si/Fe] and [Ca/Fe] observed in GC stars. However, no single
star in this mass range can simultaneously explain all halo GC heavy-element 
abundance ratios, such as [V/Fe], [Ti/Fe] and [Ni/Fe]. These require
a combination masses for the Population III stellar progenitors. 
The various observational consequences of this scenario are discussed.  

\end{abstract}

\keywords{stars: Population III---globular clusters: formation
---globular clusters: general---early universe }

\section{Introduction}

Globular clusters (GCs) are dense, bound systems 
of typically $10^4-10^6$ M$_\odot$ stars. 
They differ from galaxies in that their stellar populations are 
coeval and have extremely uniform [Fe/H].
Measurements of [Fe/H] (taken
to trace overall metallicity) show that there appear to be at least
two distinct sub-populations of GCs in the Galaxy.  One represents a
spherically distributed halo population, with mean ${\rm
[Fe/H]} \sim -1.5$, the other is spatially flattened, with 
a higher mean metallicity (${\rm [Fe/H]} \sim -0.5$) 
(Zinn 1985). This division is also seen in the 
kinematics of the two sub-populations, the halo GCs
are largely a pressure-supported system, whereas the
metal-rich GCs show signs of net rotation leading them to be
associated with the thick disk (Zinn 1985) or the bulge 
(Frenk \& White 1982).
There now exists substantial evidence for multiple populations of GCs
in external galaxies (see Ashman \& Zepf 1998 and references therein),
which exhibit similar metallicity distributions to those of the Milky Way.

The existence of correlations between the mean metallicity of the
metal-rich GCs with the mass of their parent galaxies (e.g., Forbes \&
Forte 2001), suggests that they are somehow connected to the formation
of the spheroid.
However, their halo counterparts show only weak evidence for such
correlations with their parent galaxy properties, suggesting 
different formation sites and/or formation processes for these objects.

In attempting to explain the origin of these GCs
we must address the following questions: {\it (i)} why do GCs have a
characteristic mass scale? {\it (ii)} why do they coalesce
so rapidly after the big bang (a recent mean age for Galactic GCs 
is 12.9 $\pm$ 2.9 Gyr (Carretta et al. 2000), whilst WMAP puts the age of
the Universe at 13.7 $\pm$ 0.2 Gyr (Spergel et al. 2003))?
{\it (iii)} why are their metallicities so homogeneous?  
{\it (iv)} why do they have a characteristic mean metallicity 
 of [Fe/H] $\sim$ --1.5 with a lower bound of [Fe/H] $\sim$ --2.5?  

Here we consider the view that at least some halo GCs are 
'pre-galactic' in origin (i.e., form before the bulk
of the galaxy stars).
The idea of pre-galactic GCs is not new, and was first discussed
in any detail by Peebles \& Dicke (1968). Peebles (1984) suggested
that the hierarchical clustering scenario yields 
two characteristic scales; one of which might be identified 
with GCs which form within collapsed dark matter (DM) halos
prior to galaxy formation.  Previous studies which have considered 
cooling processes in a cold dark matter (CDM) universe, 
suggest that the first stars may form within a collapsed DM 
halos of $10^5 \sim 10^7$ M$_\odot$
(e.g., Tegmark et al.\ 1997; Yoshida et al.\ 2003).
Therefore, a pre-galactic GC formation scenario can explain
the above questions {\it (i)} and {\it (ii)} naturally. 
In addition, any nucleosynthetic products, the result of star formation 
occurring in such low-mass DM halos, may be expected to have a short mixing 
timescale leading to a homogeneous chemical composition ({\it (iii)}). 
However, question {\it (iv)} is crucial. If GCs are the first objects, 
they have to form from pristine (i.e., zero metallicity) gas, and 
are therefore expected to have low metallicities (Peebles 1984). 

 In this {\it Letter}, we investigate a scenario to 
address question {\it (iv)} in particular. 
We consider that a single very massive 
($>150\Msun$) 'first star', with a consequently short lifetime 
($\sim 10^6$ years), forms in a virialised halo at very high redshift
(e.g., Abel, Bryan \& Norman 2002). The star explodes as a supernova and  
enriches the pristine gas in the halo, resulting in the formation of a GC 
with a characteristic abundance pattern. 
Recent models of the metal production and ejection
of stars with primordial compositions (Hegar \& Woosley 2002, hereafter HW02)
show that in a star of above $\sim 150\Msun$, the amount of iron produced
and ejected as a supernova jumps to several solar masses. 
This is adequate to comfortably enrich $10^6\Msun$ of gas (in a $10^7\Msun$
DM halo) to the abundances seen in Milky Way halo GCs (offering an explanation
for {\it (iv)}). 
The metals ejected from the death throws of a single giant Pop. III star
will act as an efficient coolant (Omukai 2000) significantly increasing the
local cooling rate. This may lead to a rapid burst of star
formation within enriched gas with a normal initial mass function
(IMF). Further star formation is arrested when type II supernovae 
(SNeII) within the GC expel the remaining gas, enriching and 
possibly reionising the interstellar medium (ISM) 
(e.g., Yoshii \& Arimoto 1987, Massimo 2002). 

 Before investigating this scenario in more detail, we note that 
it is quite likely that GCs form in a variety of different processes.
Indeed, there are many scenarios which address the above 
questions, and generally favour the idea that GC formation
is a purely baryonic process  (e.g. Fall \& Rees 1977; Fall \& Rees 1985; 
Caputo \& Castellani 1984; Kang et al.\ 1990; Bromm \& Clarke 2002; 
Kravtzov \& Gnedin 2003). This study does not exclude 
any of the above ideas, but we argue that on the basis of 
current theoretical understanding, and observational evidence, 
a cosmological origin for halo GCs cannot be ruled out.  
We consider that our scenario is a possible solution to the problems
alluded to previously, and here we test it severely.
Moreover, our scenario does not impinge upon the second peak 
of the GC metallicity distribution. 
Young clusters can be formed when gas clouds are
shocked, for instance during major mergers that result
in the final host. As Beasley et al.\ (2002) argued using semi-analytic
techniques, separate formation epochs and/or mechanisms may be required
to reproduce the observations. 

\section{A single very massive star}

 By definition, the first star should have primordial abundance, i.e.,
zero metallicity.  Nucleosynthesis calculations for zero metallicity
stars are still controversial, nevertheless, due to the extensive work
of several groups (e.g., HW02; Limongi \& Chieffi 2002; 
Umeda \& Nomoto 2002, hereafter UN02),
a consensus is rapidly evolving.  According to HW02, the
enriched ejecta are sensitive to the progenitor star's main
sequence mass; non-rotating stars over $260\Msun$ do not
eject any metals because they become black holes which consume
all subsequent ejecta (e.g., Fryer, Woosley \& Heger 2001).  
Between approximately 140 and $260\Msun$ lies the
domain of pair instability SNe (PISNe). These stars are completely disrupted
by nuclear-powered explosions. Below this mass
range it is again likely that black holes are the main product, and
little heavy elements are ejected (Fryer 1999; Woosley \& Weaver
1995). As PISNe do not result in a black hole, their
explosion mechanism is well understood, compared to core-collapse SNe (CCSNe),
and their ejection products
calculable. Moreover, unlike CCSNe, PISNe are not affected by
the ``mass cut'', which is the mass coordinate that separates the ejecta 
from the compact remnant, i.e., the mass inside of the mass cut has 
never been ejected.
a ``mass cut'', which is the mass coordinate at which the explosion energy is 
deposited, i.e., the mass inside of the mass cut has never been ejected
(Limongi \& Chieffi 2003).
This mass cut, which is a free parameter in CCSNe models and
to some degree governs the iron yield, is not needed in PISNe
models.

We consider stars within the mass range of 140 to
$260\Msun$ and adopt the yields for those stars calculated by
HW02.
Fig.~\ref{fehm-fig} shows the expected iron abundance when a single
massive PISNe enriches a given mass of pristine gas. The
horizontal solid line shows the observed mean metallicity
for Galactic halo GCs ([Fe/H]$=-1.62$), the dotted lines show 
the dispersion on this value ($\sigma=0.32$; calculated from the 
February 2003 version of the Harris (1996) catalogue).  
The lines for each star with
indicated masses are calculated by [Fe/H]$=\log({\rm
M_{ej,Fe}}(m_s)/(0.76 {\rm M_g})) -\log(Z_{\rm Fe,\odot}/Z_{\rm
H,\odot})$. M$_{\rm ej,Fe}(m_s)$, M$_{\rm g}$, $Z_{\rm
Fe,\odot}$, and $Z_{\rm H,\odot}$ are the iron yield for a
$m_s\Msun$ star, the mass of pristine gas, the solar metallicities of iron
and hydrogen respectively. The value of 0.76 is the 
primordial hydrogen abundance.  
We employ the solar abundance shown in Anders and Grevesse 
(1989; see Woosley \& Weaver 1995), 
and assume that the ejected metals are homogeneously
distributed, i.e., the one zone model.

The iron yields of HW02 increase with increasing
stellar mass. For a typical GC of $10^5\Msun$, forming stars at 100\%
efficency within a $10^6\Msun$ halo, the mass of a single Pop III
star must be above $180\Msun$. Since the star formation efficiency is
more likely to be $\sim$ 10\%, the mass of the first star must be 
$> 220\Msun$ embedded in a $10^7\Msun$ halo.  
This mass range is consistent with the mass of the first
stars predicted by numerical simulations ($\sim 200\Msun$; e.g., Abel,
Bryan, \& Norman 2000). Thus, Fig.~\ref{fehm-fig} demonstrates that
the explosion of a {\it single} massive star is capable of
enriching gas to the metallicity levels seen in halo GCs.
This may explain the characteristic range in the 
metallicities of halo GCs, as the individual DM halos have a
characteristic size, and the stars a limited mass range. 
Note that a mass-metallicity relation
for GCs is not necessarily expected in this model (e.g. Murray \&
Lin 1992), since there may well be dependency between M${\rm_g}$ 
and stellar mass. 


\vbox{
\begin{center}
\leavevmode 
\hbox{ \epsfxsize=8.5cm \epsffile{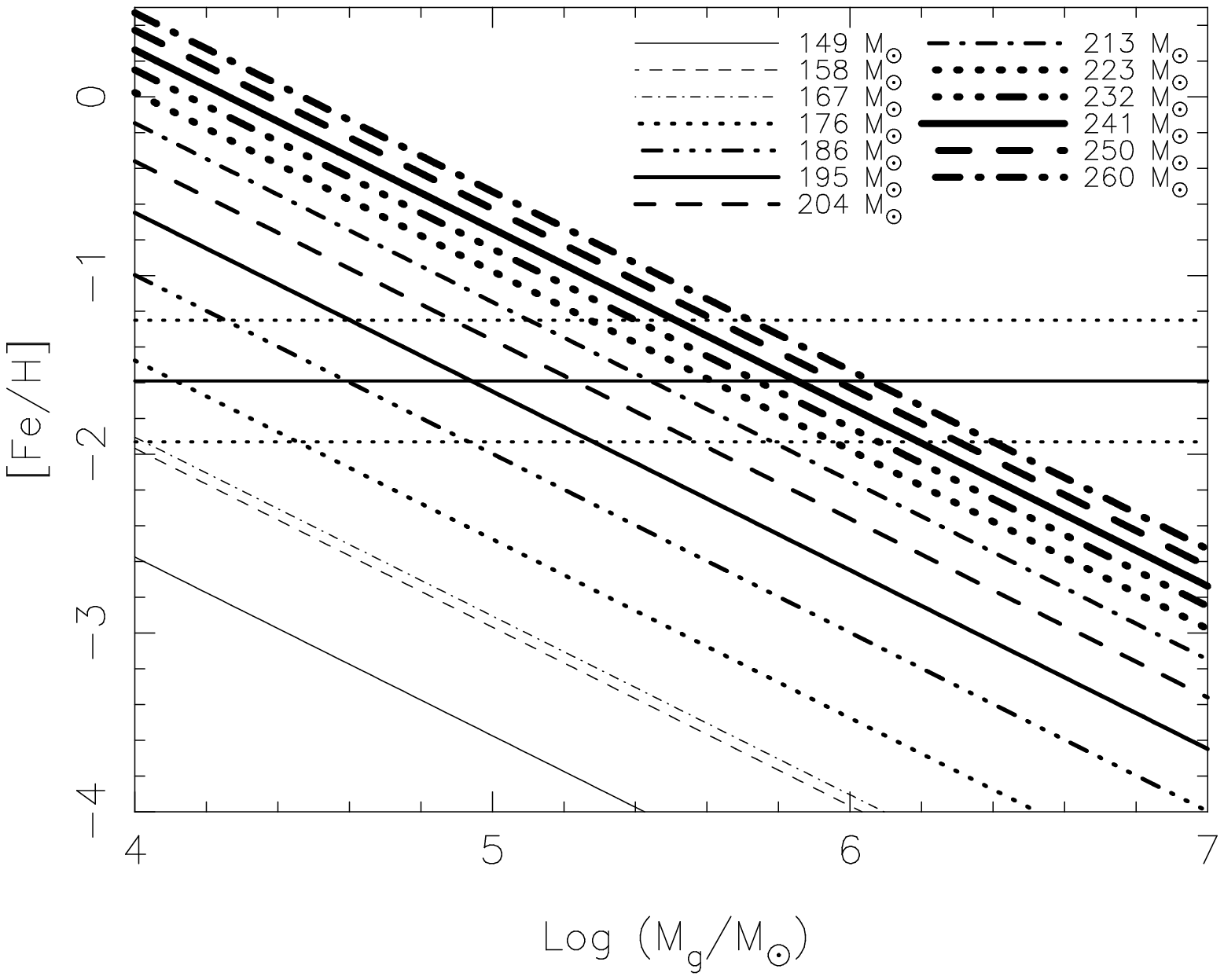}}
\figcaption[f1.eps] { 
The iron abundance produced when a single
massive PISN of a given mass enriches the amount of pristine gas shown
on the ordinate.  The horizontal line shows the observed mean
metallicity ([Fe/H]=$-1.62$) for halo GCs. Dotted lines show the
dispersion ($\sigma=0.32$; see text).
\label{fehm-fig} }
\end{center}}

 Here we consider the case that the first star is a single very massive star,
and induces a PISN. However, stars with main sequence masses 
between $\sim8$ and $\sim40$ M$_\odot$ can also produce heavy
elements through CCSNe. Although, as mentioned previously,
the iron yields of CCSNe are affected by poorly constrained parameters 
such as the mass-cut, the expected highest possible iron 
yields from CCSNe (30 M$_\odot$ star in UN02) are $\sim$ 0.2 M$_\odot$.
To enrich the halo pristine gas to the same level as a 
232 M$_\odot$ star, which produces 19.4 M$_\odot$ iron,
at least $\sim95$ CCSNe are required. Thus, although multiple CCSNe are
another possible means of adequate iron enrichment, 
a total of 2850 M$_\odot$ of gas must be converted 
to stars effectively simultaneously. 
Therefore, although not a unique solution, a single PISN is a more
efficient (and plausible) enrichment mechanism.

\section{Elemental abundances}

 One consequence of our scenario is that the observed abundances for
different metals within a single GC should be due to enrichment from a
{\it single} star. 
By comparing the observed abundances
with the predicted yields of HW02 we may read off the mass of 
the first star responsible for each GC.  
Fig.~\ref{ratio-fig} shows the various
abundance ratios of PISN yields of HW02 as a function of
progenitor stellar mass. The dotted lines are the observed range of 
the abundance ratios for stars in several Galactic GCs.
The observed abundance ratios come from high resolution
spectroscopic data of RGB stars, in which the lighter elements,
such as O, Na, Ma, and Al, are possibly affected by 
stellar evolution such as dredge up.
Thus, we here focus on relatively heavy elements, i.e., heavier
than Si, which should be unaffected by such evolutionary processes.
The panels of [Si/Fe] and [Ca/Fe] show that a star of mass 
between 220 and 240 M$_\odot$ is in remarkably good agreement with the
observational data. Such stellar masses are consistent with the range 
discussed previously. 
Thus, a star with mass around 230 M$_\odot$ is a strong candidate for the 
first star in our GC formation scenario.

\vbox{
\begin{center}
\leavevmode
\hbox{
\epsfxsize=8.5cm
\epsffile{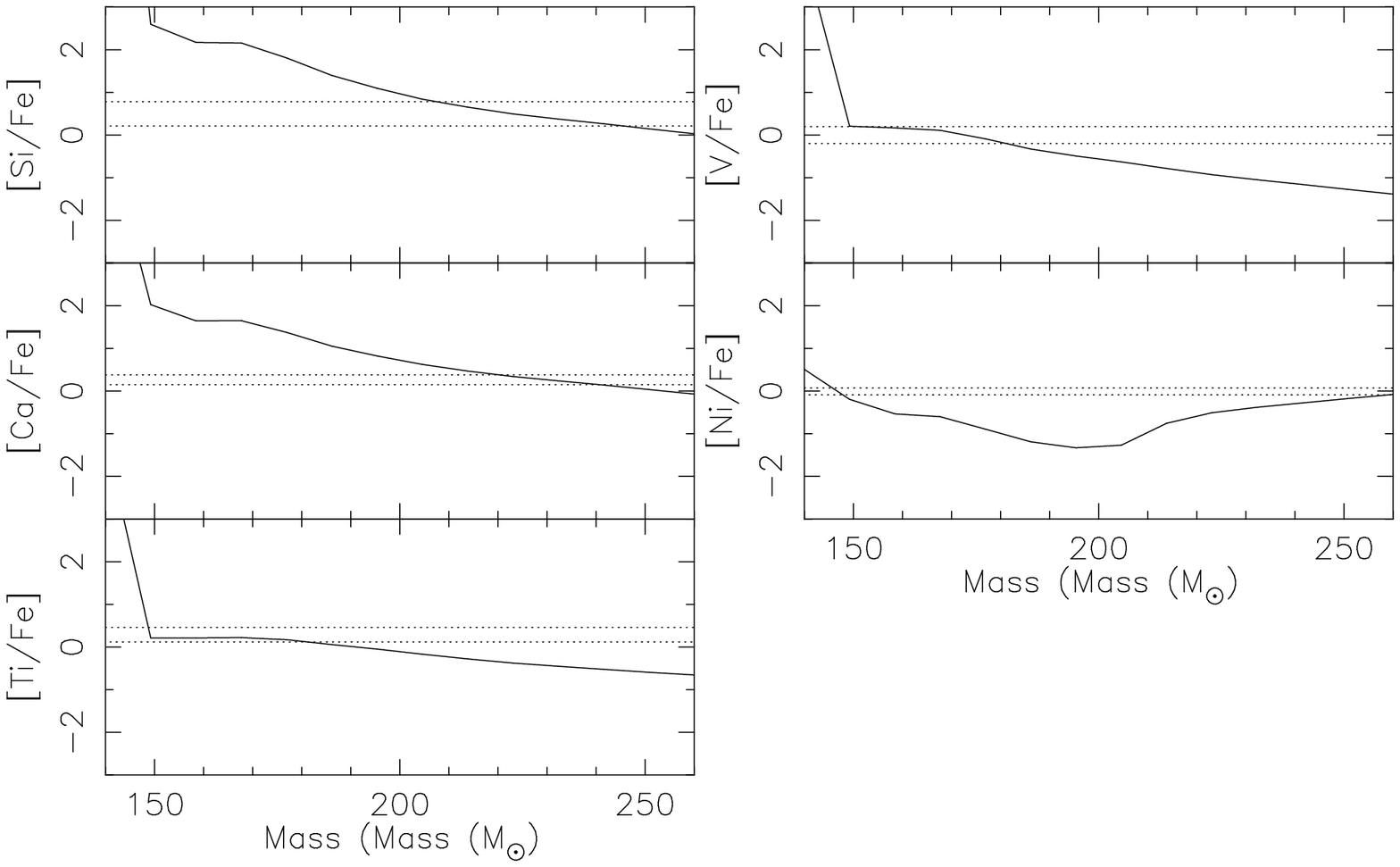}}
\figcaption[f2.eps]{ 
The abundance ratios from PISN yields as a function of the main sequence 
mass of the progenitor star. The dotted line shows the range,
i.e., maximum and minimum, of the abundance ratios in 
observed GCs (NGC 6287, NGC 6293, NGC 6541: 
Lee \& Carney 2002; M4: Ivans et al. 1999; NGC 7006: Kraft et al.\ 1998;
M15: Sneden et al.\ 1997; M13 Kraft et al.\ 1997).
\label{ratio-fig} }
\end{center}}

 However, we find that the yields for this mass cannot reproduce all 
the abundance ratios observed. 
For example, the other panels of Fig.~\ref{ratio-fig} 
demonstrate that [Ti/Fe], [V/Fe], and [Ni/Fe] for
a 230 M$_\odot$ star is significantly below the observed range.
The observed range of [Ti/Fe] and [V/Fe] requires 
a star of $<$ 180 M$_\odot$, whilst [Ni/Fe] requires 
a star with less than 150 M$_\odot$, or greater than 250 M$_\odot$.
This is clearly a problem for the model; the theoretical yields and 
observations could be reconciled by postulating 
enrichment from a small number of stars rather than a single star, e.g.,
the combination of a 230 M$_\odot$ stars, and several $<150$ M$_\odot$ stars.
However, the most serious problem of PISN enrichment for GCs is the 
absence of any elements above the iron group due to a lack of 
any s- and r-process (HW02). 
In the stars of the GCs in Fig.~\ref{ratio-fig}, 
elements such as Ba, La, and Eu are present in significant quantities.
Additional physics such as rotation of the massive
star, or subsequent pollution of the immediate environment could 
provide solutions to this problem.
Alternatively, there may  some additional contribution
from normal CCSNe, although this makes the scenario less appealing.

Since accurate stellar abundances are crucial to this 
issue, and the sample of GCs with such measurements 
is small, we strongly encourage high resolution spectroscopic
measurements of individual stars in a larger sample of GCs.
In addition, since many species may be effected by the mixing of nuclear 
products in cluster giants  (e.g. Kraft 1994), 
such analyses should be applied to main-sequence stars providing 
reliable abundance ratios for lighter elements such as [O/Fe] and [Mg/Fe].

As a result, although the proposed scenario can explain several observational
properties of GCs, there some are serious problems.
The current generation of population III yields favour
the formation of multiple stars (which should be more than 8
M$_\odot$) over a single massive star, with the inclusion
of a source of neutron-capture elements.

\section{Observational consequences}

 There are several interesting observational consequences
of our proposed scenario. If the abundance ratios predicted by the
PISN yields can be reconciled with the observations, and they are the 
result of a single massive first star,
then the mass of that star could be derived. 
With sufficiently detailed
abundance ratio observations of GCs, the mass of the precursor
Pop. III object could be extracted, which, combined with the mass
of the GC would allow the star formation efficiency to
be calculated. 

If all Pop. III stars were very massive, then none will survive to the
present day.  Very low metallicity stars, such as that recently
observed in the halo of our galaxy (Christlieb et al.\ 2002) with
${\rm [Fe/H]} \sim -5$ could in fact be {\it younger} than the halo
GC population. Such stars may simply have formed out of gas enriched by 
some secondary `normal' SNe II in the GCs, and such low  abundances could 
just be the result of dilution with pristine, or very metal-poor gas
(for an alternative view, see Limongi et al.\ 2003).

The GCs produced in our scenario form out of enriched gas such that they
have a normal IMF.  They produce normal numbers of SNeII
\footnote{The number of SNe II (from stars with the mass of $>8 {\rm M}_\odot$)
expected from a Salpeter IMF
(mass range between 0.1 and 60 M$_\odot$) is $\sim$ 730.
The total energy of SNe II (730$\times$10$^{51}$ erg)
is slightly larger than the energy of 230 M$_\odot$ PISN 
($60\times10^{51}$ erg). Subsequent to star formation, which 
consumes the gas, the remaining gas density would be low. 
Therefore, we expect that these SNeII would have more power
to blow out the remaining gas from the system than a single PISN.}
which serve to end further star formation by expelling the remaining gas, a
mechanism which naturally produces a population of stars with a very
uniform age. The gas expelled by these SNe would have abundances typical of
a more normal stellar population, rather than those for Pop. III stars. 
Thus field stars, the diffuse ISM and the
Lyman$-\alpha$ forest could consist of gas enriched by normal 
SNeII products.

Is is also interesting to speculate whether or not these
GCs would be visible using the next generation telescopes, such as the
{\it James Webb Space Telescope} (Carlberg 2002). If detectable, 
then the angular correlation function would make it possible to
distinguish between GCs in their own halos or GCs associated with
galaxies as larger halos have a higher correlation
amplitude. Although many of our halo GCs will have subsequently
fallen into larger halos (Monaco \etal 2002) some will have survived
as free floating GCs to the present day (e.g., West 1993).

\section{Discussion and Conclusions} 

 We have demonstrated that there does indeed appear to be enough iron
production within such a single massive star to enrich
$10^6\Msun$ of gas to the levels required to reproduce `halo' GC
iron abundances. Also, abundance ratios such as
[Si/Fe] and [Ca/Fe] predicted from the yields of stars with around
$230\Msun$ are consistent with those observed in Galactic
GCs. However a deficit other elements (e.g. Ti, V, N) and
complete absence of elements above the iron group 
results in inconsistencies with the current observational data.
This problem could be circumvented either by
postulating an initial giant binary system, fragments of several stars
(Nakamura \& Umemura 2001), or missing physics in the 
nucleosynthesis models of giant stars, such as rotation.  
Further investigation of this point requires
both more sophisticated numerical simulations, and 
more detailed models of high mass,
zero metallicity stars.
Irrespective of these difficulties, we argue that the detailed
abundance analysis of main-sequence stars in GCs provides
important clues to the nature of Population III objects.

We have assumed that the pristine gas in the collapsed halo
was not completely blown away by the first PISN.  
For $230\Msun$ the energy is $\sim60\times10^{51}$ ergs (HW02).
If the gas fraction in the collapsed halo is the same as the
cosmic mean ($\Omega_b/\Omega_0=0.04/0.27$: Spergel et al.\ 2003), and
both the DM halo and the gas follow the density profile
suggested by Navarro, Frenk, \& White (1997), then for a formation
redshift, $z=15$ and a concentration parameter, $c=4$, the binding
energy for a gas mass of $10^6\Msun$ is predicted to be
$\sim3\times10^{51}$ ergs.  As this is well below the hypernova
energy, we should expect the gas to be blown out of the halo,
preventing the induction of a second generation of star formation that
makes the GC.  However, according to high resolution 1-d
simulations of a single SN remnant by Thornton et al.\ (1998), 90 \%
of an initial SN energy is lost to radiative cooling in its early
expansion phase. In this case the SN energy for $230\Msun$ becomes
comparable to the binding energy. Moreover, the ambient gas density
increases with redshift, and radiative cooling is expected to be even
more efficient in such high density gas. High resolution 
numerical simulations of a single SN in a collapsed halo at
high redshift would be extremely useful to test this idea
(e.g., Mori, Ferrara, \& Madau 2002; Bromm, Yoshida, \& Hernquist 2003
\footnote{Mori et al.\ (2002) focus on the effect of multiple SNe on
a slightly larger system 10$^8$ h$^{-1}$ M$_\odot$. Bromm et al. (2003)
consider a single PISN effect on a system of a similar size to that which 
 we consider here. However, their poor DM particle resolution is 
 reason for concern, which is likely to underestimate the gravitational 
potential of DM halo.}). 
Such simulations would enable us to investigate how smoothly metals 
are mixed with halo gas, since our picture assumes an 
instantaneous, well-mixed solution.

Finally, we discuss perhaps the most important hurdle 
for any cosmological GC scenario.
Moore (1996) pointed out that the prominent tidal
tails observed for some GCs, such as M2 (Cudworth \& Hanson 1993)
suggests that such GCs do not possess DM halos.
He supported this argument with a simulation of a GC orbiting in a
static potential. The analytic argument presented by Moore (1996) derives the
tidal radius for M2 with and without DM, with the former
being closer to the observed truncation radius of 60pc. However, this
argument does not discuss the formation time of the GC or the
characteristic radius of the GC DM halo. The characteristic
radius of a $10^7\Msun$ halo forming today is several kpc, much larger
than the tidal radius derived for M2. GCs 
form at high redshift, therefore their characteristic radii are much
smaller and they could potentially have retained much of their
halos. This argument misses the major drawback of this simple analytic
approach -- the environment of the GC is not expected to remain smooth
during the formation epoch of a large host halo, which is built up by
a succession of mergers. Even today, at the radius of $8-10$ kpc, the
halo is not symmetric as the potential of the disk is not negligible,
supplying periodic impulses to any orbiting object.  We believe that
the above estimate for the tidal radius is too simplified.  
Numerical simulations similar to those of Moore (1996), 
but including the active 
evolution of the halo potential and/or the disk potential would 
be an extremely useful test for the existence of DM in GCs.



\vskip 0.5cm

MAB thanks the Swinburne RDGS.
FRP would like to thank the Royal Society for the provision of a
travel bursary to Australia during which visit this work was
completed. FRP holds a PPARC Advanced Fellowship.  DK acknowledges the
Australian Research Council through the Large Research Grant Program
(A0010517)

\end{document}